\documentclass[11pt,twoside]{article}
\usepackage{asp2004}
\usepackage{psfig}
\usepackage{epsf}
\usepackage{graphics}
\usepackage{lscape}
\def\edcomment#1{\iffalse\marginpar{\raggedright\sl#1\/}\else\relax\fi}
\reversemarginpar

\begin{document}
\title{Cosmic abundances: The impact of stellar duplicity}
\author{A. Jorissen, S. Van Eck}
\affil{Institut d'Astronomie et d'Astrophysique, Universit\'e Libre de Bruxelles, CP 226, Boulevard du Triomphe, B-1050 Bruxelles, Belgium}

\begin{abstract}
The mass-transfer
scenario links chemical peculiarities with stellar duplicity for an
increasing number of
stellar classes (classical and dwarf barium stars, subgiant
and giant CH stars, S stars without technetium, yellow symbiotic stars,  
WIRRING stars, Abell-35-like nuclei of planetary
nebulae...). Despite these successes, 
the mass-transfer scenario still faces several problems:
What is the mass-transfer mode? Why orbital elements of dwarf
barium stars do not fully match those of the classical barium stars?
What is the origin of the few non-binary stars among dwarf barium stars?
\end{abstract}
\thispagestyle{plain}

\section{The mass-transfer scenario: The origins}

The impact of the binary nature of a star on its chemical composition
seems to have been recognized first in the context of Am and Ap stars
in the years 1960--1968 
\citep{Bonsack-61,Fowler-65,Conti-65,VandenHeuvel-67a,VandenHeuvel-67b,
VandenHeuvel-68a,VandenHeuvel-68b,VandenHeuvel-68c}.
\citeauthor{VandenHeuvel-68a}\footnote{At the time, E. Van den Heuvel
 was still affiliated
with the Astronomical Institute of Brussels Free University!} (\citeyear{VandenHeuvel-68a}) 
writes: {\it[many properties of] Ap and Am
stars (...) can all be explained if these stars were originally the less
massive members in spectroscopic binaries in which the primary during
its post-main-sequence evolution filled its Roche lobe, transferred a
large fraction of its mass towards the secondary and finally evolved
into a white dwarf}. Current views about the role of duplicity among Am and Ap stars 
will be shortly reviewed in Sects.~\ref{Sect:bariumdwarfs} and \ref{Sect:filiation} 
   
Two decades later, the {\it mass-transfer scenario} popped up again,
this time in relation with barium stars and following the discovery by
\citet{McClure-80} of a large frequency of spectroscopic binaries
among the family of barium stars. According to McClure, {\it it is not unreasonable to
conclude that BaII stars are all binaries with low-mass secondaries
consisting of degenerate objects. It is possible that these systems
are such that mass has been lost from a more massive evolving star and
deposited onto the present primary, the secondary having now evolved
to the white dwarf stage.  The carbon and s-process
elements\footnote{For a review about the s-process of nucleosynthesis
and its major astrophysical site, the asymptotic giant branch (AGB)
stars, the reader is referred to \protect\citet{Lattanzio-Wood-03}} in
this case could be dumped onto the present primary-star atmosphere.
(...)  The CH stars, which are probably the Population II equivalent
of BaII stars, may also, by implication, be binaries}.

These are the first definitions of the 
{\it mass-transfer scenario} that appeared in the literature. From the
very beginning, this scenario was thus supposed to pollute not only giant
companions but also dwarf ones. 

\section{The early problems with the mass-transfer scenario: 
David Lambert's question {\it Where are the barium dwarfs?}}
\label{Sect:bariumdwarfs}

Although willing to support the mass-transfer scenario, David Lambert
nevertheless raised the question {\it Where are the main-sequence
progenitors of the classical Ba stars?} in his review talk
\citep{Lambert-88} at the IAU Symposium 132 in Paris where one of the
authors (AJ) first met him.  This question naturally arose from the
realization that the mass-transfer scenario does not pollute more
efficiently giant companions than dwarf companions.  Since the
accretion flow is governed by gravity and by pressure forces, the
actual accretion cross section is much larger than the star
geometrical cross section
\citep{Bondi-Hoyle-44,Boffin-Jorissen-88,Theuns-96}. Therefore, dwarf
stars polluted by mass transfer should be at least as common among
main sequence stars as barium stars are among G-K giants \citep[a few
percents;][]{McClure-1984:a}.  Yet, apart from the carbon
dwarf G77-61 \citep{Dahn-1977:a}, the only polluted dwarf stars
identified at the time were the CH subgiants \citep{Bond-74}, and
those had discrepant Li abundances as compared to the barium giants
\citep[as well as discrepant neutron exposures;][and Busso
et al. 2001 for an update on this question]{Lambert-88}. Since barium
dwarfs should be found all along the main sequence, it was thus puzzling
that subgiant CH stars were restricted to a very narrow temperature range
around G0
\citep{Bond-74,Lambert-88}.

These objections were lifted several years later, when
\citet{Lambert-93} realized that the resonant Li~I line in the barium
giants was blended, so that earlier Li abundances were
overestimated. Classical barium stars and subgiant CH stars now had
abundances consistent with their suspected evolutionary link
\citep[see also][]{Smith-93}.

As for the second objection, \citet{Tomkin-89} quite appropriately
remarked that {\it among the stars of spectral types A to F, discovery
of the barium stars may have been hampered by the presence of large
numbers of chemically peculiar stars. Indeed, the diffusive separation
that leads to abundance anomalies in the atmospheres of the normal
stars may operate too in the young barium stars so obscuring their
true nature}. The bright F5V star HR~107 was the first example
\citep{Tomkin-89} of a candidate barium dwarf warmer than the subgiant CH
stars (later studies did not confirm the binary nature of this
star, though; see Sect.~\protect\ref{Sect:Non-binarydwarfs}). A few
years later, \citet{North-Duquennoy-91} identified many more dwarf
barium stars among the stars flagged by Bidelman as F str $\lambda$4077 in
the Michigan spectral survey. Nowadays, several cool C dwarf stars
\citep[collected in Table 9.5 of][]{Jorissen-03}, blue metal-poor
stars \citep{Preston-Sneden-00,Preston-Sneden-01} and `Wind-Induced
Rapidly Rotating' K dwarfs \citep[known as WIRRING stars
after][]{Jeffries-Smalley-96,Jeffries-Stevens-96} have joined the
family of barium dwarfs. The possible incidence of the mass-transfer
scenario among Am stars is not quite clear yet
(Sect.~\protect\ref{Sect:filiation}). Furthermore, a definite proof
of the filiation between these various classes requires as well a careful
comparison of their respective chemical
composition.\footnote{Both  \citet{Lambert-85}  and
  \protect\citet{North-94} remark for example that the abundance pattern of Am and Fm
  stars differ markedly from those of barium and F str $\lambda$4077 stars, but
remember the quotation from \citet{Tomkin-89} earlier in this section} Such a comparison
is beyond the scope of the present review, which is restricted to a discussion
of the orbital elements. The reader interested in a comparison of the
chemical compositions is referred to
\citet{Lambert-85,North-94,Busso-01}. 

\section{Successes of the mass transfer scenario} 
\label{Sect:success}

\begin{figure}
\plotone{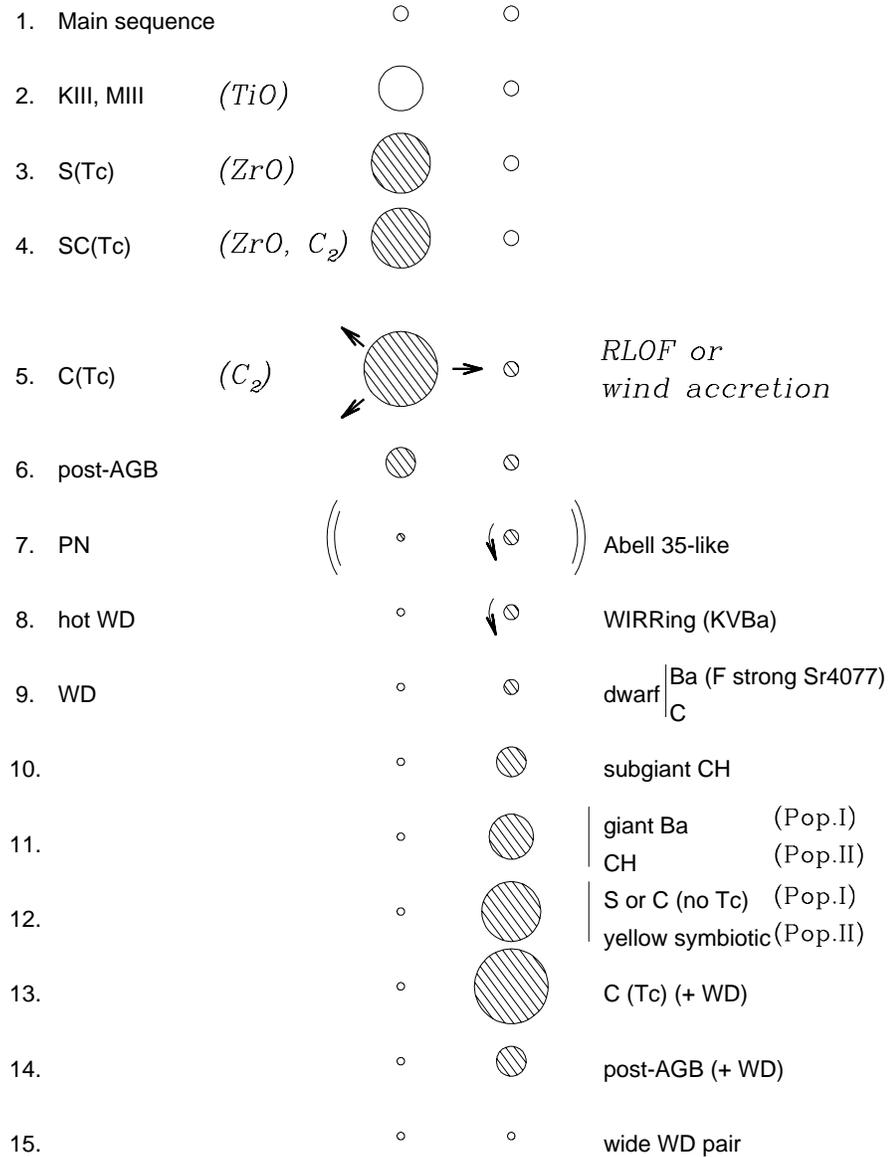}
\caption[]{\label{Fig:binaryscenario}
The evolution of a
system consisting initially of two low- or intermediate-mass
main-sequence stars. The left column corresponds to the normal
evolutionary sequence of single stars, while the right column
represents the various classes of stars with chemical peculiarities 
specifically produced by mass transfer across the binary system. Hatched circles denote 
stars with atmospheres enriched in carbon or heavy elements
\citep[see]
[for a detailed description of the various stellar families involved]{Jorissen-03}.
}
\end{figure}

The mass-transfer scenario now 
encompasses many more stellar families than in the early days, as
displayed in Fig.~\ref{Fig:binaryscenario}. This growth is in itself an illustration of the success 
achieved by this scenario; more specifically  \citep[see][for a more
detailed review]{Jorissen-03}:
\begin{itemize}
\item The filiation barium stars -- S stars without technetium [`S(no-Tc)'] is fully endorsed by the similarity of their orbital elements;
\item Yellow symbiotic stars have been shown to be the low-metallicity, high-luminosity counterparts of barium stars \citep[the CH giants mentioned in Sect.~1 have lower luminosities than yellow symbiotics; see Fig.~3 of ][]{Jorissen-03a};
\item WIRRING stars and Abell-35-like binary nuclei of planetary nebulae all show, on top of their s-process overabundances, signatures of rapid rotation, most probably caused by spin accretion accompanying mass accretion, as predicted by hydrodynamical simulations of mass
transfer.   
\end{itemize}

\section{Recent problems with the  mass-transfer scenario}  

Despite the successes sketched in Sect.~\ref{Sect:success}, 
the mass-transfer
scenario is currently not without problems. First, several barium dwarfs do not seem to be binaries
(Sect.~\ref{Sect:Non-binarydwarfs}). Second,  the exact nature of the mass
transfer process is not fully elucidated yet
(Sect.~\ref{Sect:mass-transfer-mode}).  Third, the expected filiation
between post-AGB stars, barium dwarfs and barium giants is not
fully supported by  their eccentricity
-- period diagrams (Sect.~\ref{Sect:filiation}), since there is an excess of short-period,
large-eccentricity systems among post-AGB and dwarf barium stars. 

\subsection{What is the nature of the non-binary  barium dwarfs?}
\label{Sect:Non-binarydwarfs}

The family of dwarf stars enriched in carbon and s-process elements
appears to be a very heterogeneous one, as can be seen on
Table~\ref{Tab:dwarfBa}.  For instance, among subgiant CH stars, `disk' 
stars behave differently from low-metallicity stars as far as their
binary frequency is concerned \citep{Preston-Sneden-01}. Similarly, among
the 6 `mild' Ba dwarfs   ($0.23 \le [{\rm  Ba/Fe}] \le 0.68$)  found from
a high-accuracy abundance survey of 200 F dwarfs in the solar
neighbourhood \citep{Edvardsson-93}, only HR~4395 seems to be a binary.
On the contrary, the frequency of binaries is close to 100\% among `strong' Ba dwarfs
(with $[{\rm  s/Fe}]
\ga 0.5$), consisting mainly of the ``FV str $\lambda 4077$'' stars and
the disk CH subgiants. It is also suspected to be 100\% among the 
blue metal-poor stars with Ba overabundances.

Table~\ref{Tab:dwarfBa} thus reveals that, among Ba dwarfs, the 
binary frequency  is either consistent with 100\% or close to 0\%.
The existence of non-binary, low-luminosity stars with Ba
overabundances immediately raises the question of the origin of their
chemical anomalies. Three broad classes of scenarios have been proposed
so far: (i) `non-standard' stellar evolution; (ii) disguised mass-transfer
scenarios; (iii) coalescence of the two components of a binary system, 
or even possibly planet-engulfing.

\begin{table}[t]
\caption[]{\label{Tab:dwarfBa}
The various families of dwarf Ba and C stars, along with the observed fraction of binaries among them.
The column labelled Refs. provides the references where
the binary statistics and the orbital elements may be
found.}
\begin{center}
\begin{tabular}{llc}
    \hline
    \hline
Family & Fraction & Refs. \cr
       & of binaries &\cr
\hline
dwarf Ba & & \cr
\hspace*{3mm}``FV str $\lambda 4077$'' and disk CH subgiants& 28/30 & [1] \cr
\hspace*{3mm}blue metal-poor with Ba overabundances & 100\%? & [2] \cr
\hspace*{3mm}Ba dwarfs among FV from [3]  & 1/6 & [4] \cr
\hspace*{3mm}low-metallicity CH subgiants& 0/3 & [5]  \cr
dwarf C & $\ge$4/31 & [6] \cr
\hline
\hline
\end{tabular}
\end{center}
 
{\small References:\\
{[}1] \citet{McClure-1997:a,North-92,North00,Preston-Sneden-01} \\
{[}2] \citet{Preston-Sneden-00,Sneden-03a,Masseron-04}\\
{[}3] \citet{Edvardsson-93}. The Ba dwarfs
are HR\,107, HR\,2906, HR\,4285, HR\,4395 (binary), HR\,5338,
HD\,6434. Their Ba overabundances (in
the range $0.23 \le [{\rm  Ba/Fe}] \le 0.68$)  
cannot be ascribed to the normal galactic chemical evolution.\\
{[}4] \citet{Jorissen-Boffin-92,North-92,North00}\\
{[}5] \citet{Preston-Sneden-01}\\
{[}6] see Table 9.5 of \citet{Jorissen-03}
}
\end{table}

The original suggestion by \citet{Bond-74} of a mixing 
that returns a post-He flash star to the subgiant region belongs to
the first category. \citet{Smith-Demarque-80} concluded, however,
that this hypothesis could work only if 
some new physics were acting. As far as low-mass, extremely metal-poor
stars are concerned, \citet{Fujimoto-00} found that they can evolve
into carbon stars at the end of their red giant branch evolution. 
As for the second class of scenarios, \citet{Beveridge-Sneden-94}
\citep[see also ][]{Lambert-01} suggest that Ba dwarfs (especially
low-metallicity ones) may form after their atmosphere has been polluted 
by pockets of ISM material enriched in carbon and s-process elements by mass loss 
from a nearby AGB star. 
The third class of scenarios (coalescence or planet-engulfing) involves a modification of the 
internal angular-momentum distribution, which may possibly lead to some non-standard 
mixing 
and nucleosynthesis, or to flare-related nucleosynthesis if the star has been 
spun up substantially.
No  studies specific to the problem under consideration have been performed yet, however
\citep[see][for a more general discussion]{McClure-97a,Siess-Livio-99}.   

 \subsection{The mass transfer mode: Importance of periastron mass transfer?}
\label{Sect:mass-transfer-mode}

\begin{figure} 
\plotone{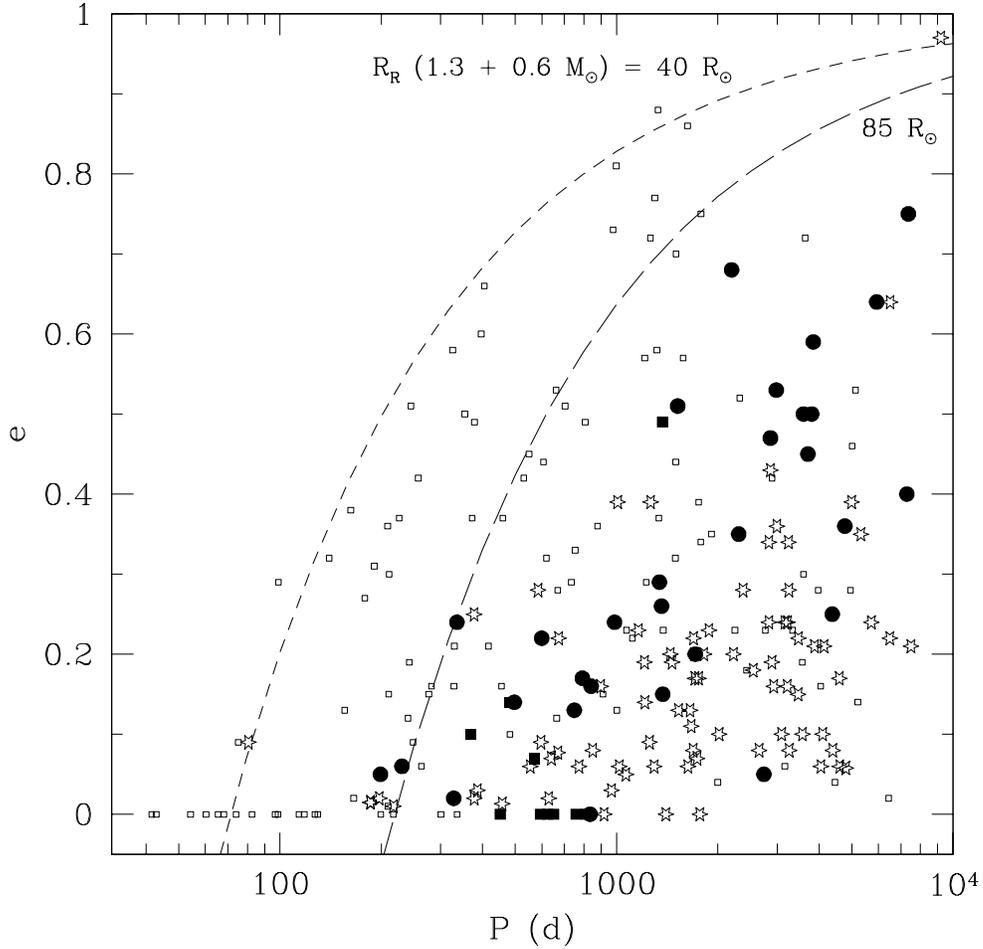}
\caption[]{\label{Fig:elogP} The period -- eccentricity diagram for
binary systems involving a KIII primary (open squares), MIII primary
(black dots; black squares for M giants in symbiotic systems) and Ba or S(no-Tc) primary (starred
symbols). The dashed lines correspond to the loci of constant periastron distance (100 and
190~$R_\odot$), translating into Roche radii of 45 and
85~$R_\odot$, respectively (assuming masses of 1.3 and 0.6 $M_\odot$
for the two components).  }
\end{figure}

\citet{Pols-03} have convincingly shown that the current prescriptions
for binary evolution (both wind accretion and Roche lobe overflow --
RLOF) fail to reproduce the orbital elements observed in barium
stars. The orbital periods of the barium systems predicted by the
synthetic binary evolution codes are either too short (when they
result from RLOF) or too long (when they result from wind accretion in
a system which must remain detached). Almost no systems are predicted
in the observed period range for barium systems ($100 \le P({\rm d})
\le 10^4$; see Fig.~\ref{Fig:elogP}). It is likely that something is wrong with our understanding of RLOF when it involves a giant star with a deep
convective envelope \citep{Iben-00,Jorissen-03}, as it is the case for
the former AGB companion which polluted the barium star.  It is generally
believed that RLOF is dynamically unstable when the mass-losing star has a deep convective envelope and is  the more massive component of the system. This situation leads to a common envelope stage and subsequently to dramatic
orbital shrinkage as orbital energy is used to expell the common
envelope \citep{Jorissen-03}.

As binary stars evolve along the sequence displayed in
Fig.~\ref{Fig:binaryscenario}, their orbital elements are expected to
vary, be it due to tidal effects, to interaction with a circumbinary
disk, or to mass transfer. Therefore, the comparison of orbital
elements for binary stars located at different stages in the sequence
is expected to shed light on this puzzle about the mass transfer mode.

Extensive sets of orbital elements are now available for binary stars
with a KIII primary \citep{Mermilliod-1996}, MIII primary
\citep[][for M giants in classical symbiotic systems, excluding symbiotic and 
recurrent novae]{Jorissen-2004:b,Mikolajewska-03}, and barium or S(no-Tc)
primary  \citep{Jorissen-VE-98}.

Figure~\ref{Fig:elogP} reveals a smooth evolution along the sequence KIII--MIII--Ba/S(no-Tc), in
the sense that the upper boundary of the populated region in the eccentricity -- period  diagram
moves towards larger periods. This is clearly a consequence of the larger radii
reached by stars evolving along this sequence. Equating the stellar radius to the
Roche radius results in a threshold period (for given component masses) below which the primary star
undergoes RLOF. Adopting the usual expression for the Roche radius $R_{R,1}$ around star 1
\begin{equation}
R_{R,1}/A = 0.38 + 0.2 \log q \:\;\;\;\;\;\;(0.5 \le q = M_1/M_2 \le
20),
\end{equation}
where $A$ is the orbital separation,  yields from Kepler's third law, $P = 70$~d for a star
of radius 40~R$_\odot$ filling its Roche lobe (adopting $M_1 = 1.3$ and $M_2 =
0.6$~M$_\odot$). Although the Roche lobe
concept is in principle only applicable to circular orbits, one may
formally compute the orbital periods for which the primary star fills
its Roche lobe {\it at periastron}, by replacing $A$ by $A(1-e)$ in the
above expression (with $e$ being the orbital eccentricity). It is
quite remarkable that the relationship between $P$ and $e$ so obtained
(assuming $R_R = 40$~R$_\odot$) exactly matches the boundary of the
region occupied by KIII giants in the $(e, \log P)$ diagram. 
This excellent match thus clearly suggests that tidal effects and/or mass
transfer at periastron play a crucial role in shaping the
eccentricity -- period diagram, through the processes described by \citet{Duquennoy-92,Soker00}.

The value
of 40~R$_\odot$ falls in the upper range of radii measured by \citet{VanBelle-99}
for KIII stars. As far as MIII binaries are concerned, a periastron Roche radius of 85~R$_\odot$
encompasses all the binaries with MIII primaries, though the match here is not as good
as for KIII binaries. This may probably be explained by the smaller
sample size, and by the larger detection biases \citep[eccentric
systems are not so easily detected among M giants, because they have
less conspicuous velocity variations, which are often confused by
their intrinsic pulsational jitter; see][]{Jorissen-2004:b}. Again, the  value of 85~R$_\odot$
derived from the $(e, \log P)$ diagram is somewhat larger than the radii measured by
\citet{VanBelle-99} for early MIII stars. This suggests that the upper left envelope observed in the
$(e, \log P)$ diagram is mainly shaped by tidal effects, which operate already {\it  before} the star
fills its Roche lobe \citep[e.g.,][]{Duquennoy-92}.

Another striking  difference between the K and M giants on Fig.~\ref{Fig:elogP} is the {\bf 
lack, among binaries involving M giants, of the many circular systems at the short-period end as
observed among the K giants}. This difference is surprising, since both stellar families involve
stars with deep convective envelopes which should react similarly to tidal effects and mass transfer.
M giants suffer, however, from a much more severe wind mass loss than K
giants, and this offers a clue to explain the differences observed in
their eccentricity -- period diagram.

Finally, it must be remarked that barium systems almost exclusively
fall in a gap observed at $P > 400$~d, $e < 0.1$ in the $(e, \log P)$
diagram of {\it pre}-mass-transfer binaries. This gap is especially
apparent in the $(e, \log P)$ diagram of G and K dwarfs of the solar
neighbourhood \citep{Duquennoy-Mayor-91}. Figure~1 of \citet{North00} 
clearly shows that barium dwarfs fall almost exclusively within this gap. 
The few KIII and
MIII binaries located in this gap
could in fact be {\it post}-mass-transfer binaries, a possibility that
would be worth testing by looking for Ba-like abundance anomalies in
those stars.

\subsection{The filiation post-AGB -- dwarf Ba -- Ba/S(no Tc) stars}
\label{Sect:filiation}

With many orbital elements now available for post-AGB stars 
 and dwarf barium stars,
it is interesting to check whether orbital elements support the
filiation post-AGB -- dwarf Ba -- Ba/S(no-Tc) stars. 

\begin{figure} 
\plotone{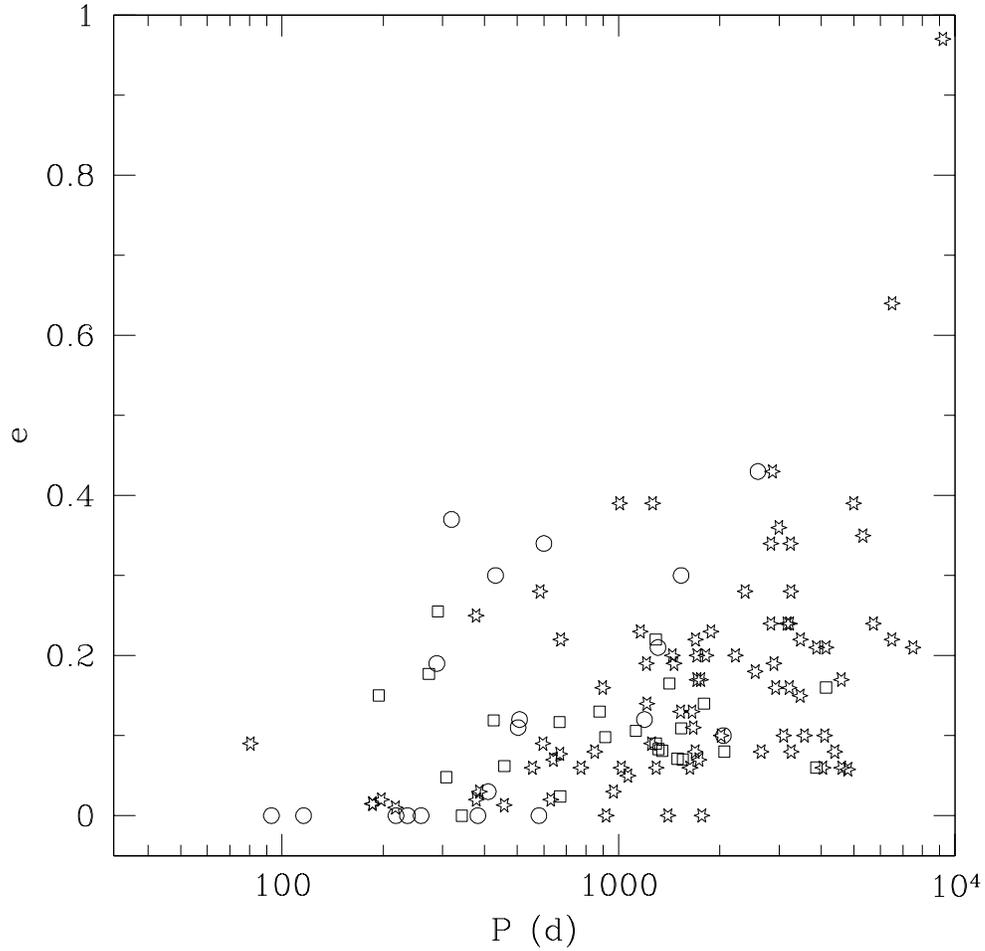}
\caption[]{\label{Fig:Badwarfs} The period -- eccentricity diagram for
binary systems involving post-AGB stars \citep[open circles, from][]{Maas-03},
dwarf barium stars
\citep[open squares,][and priv. comm.]{McClure-1997:a,Sneden-03a,North00}
and Ba/S(no-Tc) stars \citep[starred symbols, from][]{Jorissen-VE-98}.
Note that the triple system BD+38$^\circ$118 (Ba star) has not been
represented.  }
\end{figure}

Post-AGB stars and dwarf Ba stars share the same region of the
eccentricity -- period diagram (Fig.~\ref{Fig:Badwarfs}). The discrepancies between two other
properties of these families  might, however, 
 lead one to believe that binary post-AGB stars are unlikely to become Ba stars.
These (apparently) discrepant properties are the fact that (i) the mass functions of post-AGB systems
\citep[0.14--0.97~M$_\odot$;][]{Maas-03} are much larger than those of barium dwarfs
\citep[0.080~M$_\odot$;][]{Jorissen-03}, and (ii) the fact that the binary post-AGB stars do not
exhibit s-process overabundances \citep{VanWinckel-03}.

These discrepancies appear instead quite natural when one realizes that it is the {\it companion} of
the post-AGB star, rather than the post-AGB star itself, which should become the Ba star
(for post-AGB stars with main sequence companions)!
First, the chemical composition of this companion (the future Ba star) cannot unfortunately be derived
from spectral analysis, given its faintness, but it might well turn out to be rather different from
that of the current post-AGB star composition. The composition of the companion is determined by the
mass accreted in a former state of the binary system, when the post-AGB star was still on the AGB. In
the current state of the binary system, the post-AGB star is known to undergo very specific chemical
fractionation processes \citep[like gas-grain segregation, and re-accretion of gas
depleted in refractory elements;][]{VanWinckel-03}, but these processes were not necessarily operating
at the time when the bulk of the mass was accreted by the Ba star in the making.
Second, if dwarf barium stars are indeed the progeny of post-AGB stars, 
their mass functions $f_{\rm pAGB}, f_{\rm Ba}$  must be related by the simple relation $f_{\rm Ba} = f_{\rm pAGB} (M_{\rm pAGB}/M_{\rm Ba})^3$,
resulting from the fact that the (observed)  primary 
components are reversed between those two classes.
To be more specific, if $M_{\rm pAGB} = 0.67$~M$_\odot$, $M_{\rm Ba} = 1.25$~M$_\odot$
\citep[][Table~9.8]{Jorissen-03}, and  $f_{\rm pAGB} = 0.5$~M$_\odot$, then  $f_{\rm Ba} =
0.08$~M$_\odot$, in perfect agreement with the average value for Ba dwarfs \citep{Jorissen-03}!

Comparing now barium dwarfs to the {\it giant} barium and S(no-Tc) stars, there is a
significant excess of eccentric systems in the 300 -- 600~d range among
barium dwarfs and post-AGB stars. These rather short-period, eccentric
systems seem to form a distinct group from the bulk of the sample. Two
giant Ba stars (HD~58368 and HD~199939) and one S(no-Tc) star
(HD~191589) belong as well to this group of short-period, eccentric
systems. Interestingly enough, the S star HD~191589 has an unusually
large mass function of 0.394~M$_\odot$, pointing to a main-sequence
rather than white-dwarf companion. The F0-F2 main-sequence companion
of HD~191589 has indeed been detected by the {\it International
Ultraviolet Explorer} \citep{Ake-Johnson-92c}. Could it be that the
small group of short-period, eccentric systems are in fact
pre-mass-transfer systems? But how to explain then the origin of their
chemical peculiarities? These systems would then join the group of
non-binary barium dwarfs (Sect.~\ref{Sect:Non-binarydwarfs}), in the
sense that they call as well for an alternative
to the mass-transfer scenario in order to explain their chemical peculiarities. 
 
\begin{figure} 
\plotone{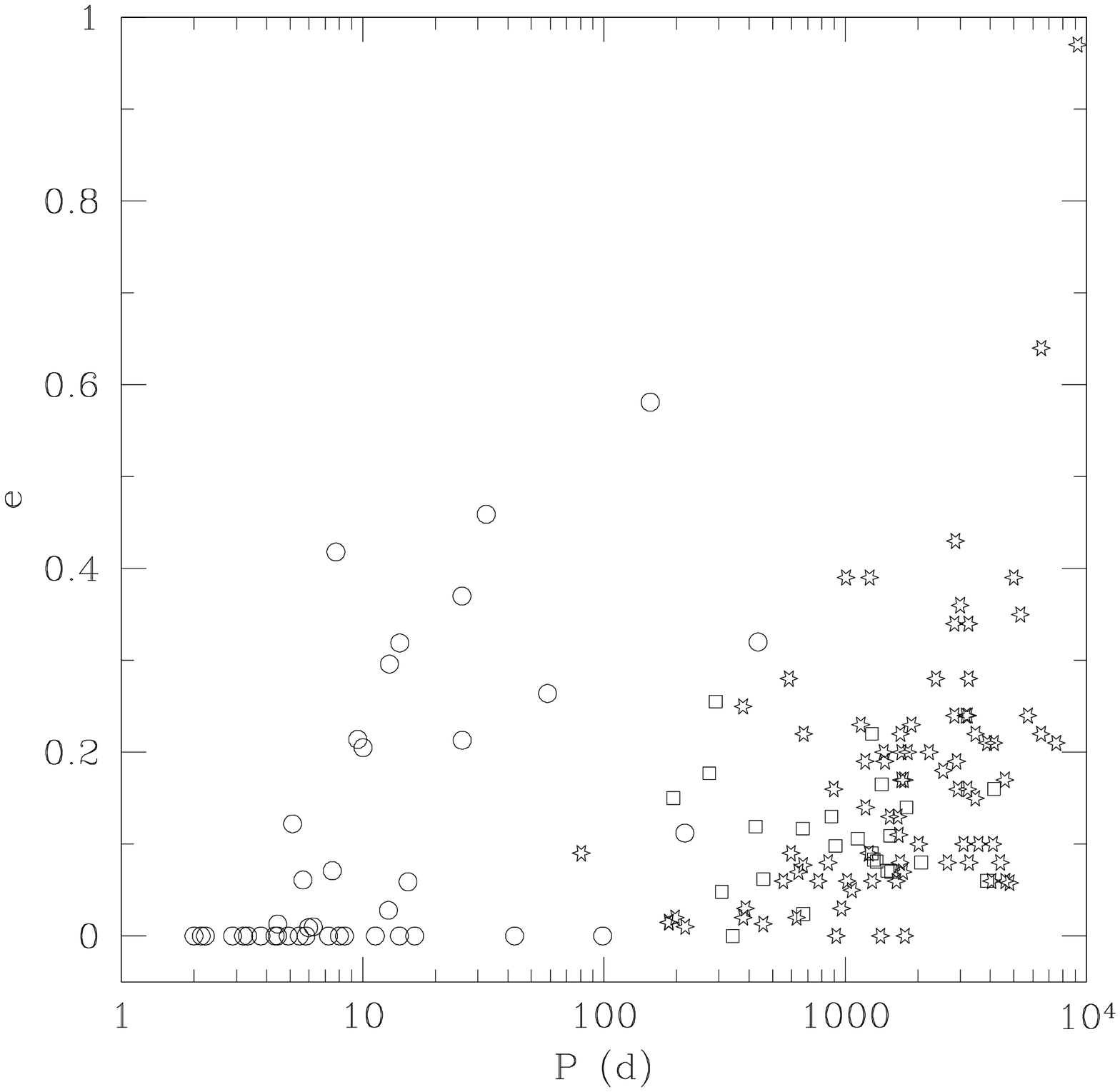}
\caption[]{\label{Fig:Am} The period -- eccentricity diagram for
binary systems involving Am stars (open circles) with orbital elements
collected from the literature, but restricted to modern data obtained
from the Swiss or Cambridge correlation spectrovelocimeters (The
long-period pairs of triple systems have not been displayed). For
comparison, Ba dwarfs and CH subgiants (open squares), and Ba/S(no-Tc)
(starred symbols) are displayed as well.  }
\end{figure}
  
To conclude this section on eccentricity -- period diagrams, 
it must be remarked that the available orbital elements for Am stars 
\citep[derived from modern spectrovelocimeter data like
  CORAVEL or ELODIE;][]{Baranne-79,Baranne-96}
do not convincingly point towards a strong 
incidence of the mass-transfer scenario among this family. As can be seen
on Fig.~\ref{Fig:Am}, the large number
of short-period ($P \la 100$~d) Am binaries contrasts with the
situation prevailing for barium stars (see Fig.~\ref{Fig:elogP}). Tidal
effects in these close binaries play a key role in synchronizing the
stellar spin with the orbital motion. Small
rotational velocities are needed for the gravitational settling to operate
and to lead to the observed chemical peculiarities \citep{Michaud-02}. A
few binaries\footnote{excluding the wide pairs in triple systems}  with periods
in excess of 100~d [HD 36360:
\citet{Carquillat-04};  KW 538, vB 130: 
\citet{Debernardi-00a}]
are present among Am stars, but they lie
at the short-period edge of the region occupied by the barium dwarfs. The conclusion that very
few -- if any -- Am binaries fall among the region occupied by Ba giants in the $(e, \log P)$ diagram must
however await confirmation from the forthcoming release of complete samples of orbital elements by
Carquillat, Debernardi and North. In the meantime, the reader is referred to
\citet{North-Duquennoy-91,North-98,Debernardi-00,Bidelman-02,Carquillat-03,Carquillat-04} for the
most recent discussions on the role of duplicity among Am and Ap stars.

\section{Conclusion}

The impact of stellar duplicity on cosmic abundances has been
investigated   for more than 35 years now.
The mass-transfer scenario, producing polluted stars
with abundance patterns otherwise incompatible with their evolutionary
stage, is widely accepted as the cause of the chemical peculiarities
exhibited by classical barium giants, S(no-Tc) stars,
CH stars, yellow symbiotics, WIRRING stars, Abell-35-like binary nuclei of
planetary nebulae... Yet this is not the end of the story, as the
mass-transfer scenario still faces some difficulties:

\begin{itemize}
\item Precisely how the mass transfer occurs is not at all understood.
Physical insight and constraints come essentially from the comparison 
of orbital elements of  pre- and post-mass-transfer 
stellar families. Such a comparison reveals 
in particular that 
(i) dynamical effects occuring at periastron are probably
decisive for the circularization and/or spiral-in of the binary system;
(ii) the total lack of short-period, circular systems among
pre-mass-transfer M giants (whereas such systems are numerous among K
giants) may hint at the key role of wind mass loss from the mass donor.

\item Orbital elements do not contradict 
the filiation post-AGB -- dwarf barium stars. The very specific
abundance
pattern of post-AGB stars seems to  result from chemical fractionation
processes occuring {\it after} the bulk of the AGB envelope has been
transferred to the companion star (the barium star in the making).
Abundance analyses of post-AGB companions are needed to confirm that
their abundances are consistent with those of barium stars.

\item The barium dwarf -- barium giant filiation is generally  well
supported  by the comparison 
of their orbital elements, with the exception, however,  of a small
excess of short-period, eccentric orbits among dwarf barium (and
post-AGB) stars. These systems might disappear through spiral-in when
they reach the red giant branch. Alternatively, these 
barium dwarfs could have a main sequence companion (rather than the
expected white dwarf), as is the case, incidentally, for the Tc-poor S
star HD~191589 which falls in the same region of the period --
eccentricity diagram. But their chemical peculiarities cannot be
attributed then to the mass-transfer scenario. These systems would thus
join the small set of Tc-poor S stars with a main-sequence companion (like
HD~191589) and of {\it non-binary} dwarf barium stars and subgiant CH
stars, which require as well an alternative to the mass-transfer scenario.


\end{itemize}

\begin{acknowledgements}
AJ first wants to thank D. Lambert for the opportunity he offered him to spend 
nine months in Texas, back in 1990. These months were a source of inspiration for
many years thereafter.    
We thank P. North for communicating us orbital
elements of barium dwarfs in advance of
publication. Discussions with A. Frankowski and J.M. Carquillat also helped to clarify 
some of the topics discussed here. AJ is Senior Research Associate from the {\it Fonds
National de la Recherche Scientifique}, SVE is Research Associate from
the {\it Fonds National de la Recherche Scientifique}.
\end{acknowledgements}

\bibliographystyle{apj} 
 
\bibliography{ajorisse_articles}

\noindent {\bf W. P. Bidelman:} I have recently decided that the most
interesting binaries are those that have coalesced and become single
stars, though we may not know it. I have thought the Ap stars might be
such objects.
\end{document}